\documentclass[12pt]{article}
\usepackage{graphicx}
\usepackage{amsmath}
\usepackage{amssymb}
\usepackage{tikz}
\usepackage{hyperref}

\numberwithin{equation}{section}

\def\bC{\mathbb{C}}
\def\bH{\mathbb{H}}
\def\bR{\mathbb{R}}
\def\bZ{\mathbb{Z}}

\def\cN{\mathcal{N}}
\def\fg{{\mathfrak{g}}}
\def\fh{{\mathfrak{h}}}
\def\SL{\mathrm{SL}}
\def\SU{\mathrm{SU}}
\def\SO{\mathrm{SO}}
\def\USp{\mathrm{USp}}
\def\U{\mathrm{U}}
\def\Tr{\mathrm{Tr}}
\def\Aut{\mathrm{Aut}}
\def\Out{\mathrm{Out}}
\def\Inn{\mathrm{Inn}}
\def\rank{\mathrm{rank}}
\def\diag{\mathop{\mathrm{diag}}}
\def\vev#1{\langle#1\rangle}
\def\Re{\mathop{\mathrm{Re}}}
\def\Im{\mathop{\mathrm{Im}}}
\def\top#1{\widetilde{\mathrm{O}#1}{}^+}
\def\aff#1{\overset{\circ}{#1}{}}
\def\KO{KO}
\def\KSp{K\!Sp}

\def\node#1#2{\overset{#1}{\underset{#2}{\circ}}}
\def\ver#1#2{\overset{{\llap{$\scriptstyle#1$}\displaystyle\circ{\rlap{$\scriptstyle#2$}}}}{\scriptstyle\vert}}

\usetikzlibrary{arrows}
\tikzstyle{every picture}+=[remember picture]
\tikzstyle{na} = [baseline=-.5ex]
\tikzstyle{mine}= [arrows={angle 90}-{angle 90},thick]

\def\Llleftarrow{%
\lower2pt\hbox{\begingroup
\tikz
\draw[shorten >=0pt,shorten <=0pt] (0,3pt) -- ++(-1em,0) (0,1pt) -- ++(-1em-1pt,0) (0,-1pt) -- ++(-1em-1pt,0) (0,-3pt) -- ++(-1em,0) (-1em+1pt,5pt) to[out=-105,in=45] (-1em-2pt,0) to[out=-45,in=105] (-1em+1pt,-5pt);
\endgroup}
}

\begin{document}

\begin{titlepage}

\begin{flushright}
IPMU-11-0166
\end{flushright}
\vskip 3cm
\begin{center}
\def\thefootnote{\fnsymbol{footnote}}
{\Large \bfseries
On S-duality of 5d super Yang-Mills  on $S^1$
}

\vskip 1.2cm

Yuji Tachikawa

\bigskip

IPMU, University of Tokyo, Kashiwa, Chiba 277-8583, Japan

\vskip 1.5cm

\textbf{abstract}

\bigskip

\end{center}
{
We study a duality of 5d maximally supersymmetric Yang-Mills on $S^1$,
which exchanges the tower of Kaluza-Klein W-bosons and the tower of instantonic monopoles. 
This duality maps a non-simply-laced gauge theory to a simply-laced gauge theory twisted by an outer automorphism around $S^1$, and is closely related to the Langlands dual of affine Lie algebras. 

We also discuss how this S-duality is implemented in terms of 6d $\cN=(2,0)$ theory.
This is straightforward except for the 6d theory of type $\SU(2n+1)$ with $\bZ_2$ outer-automorphism twist, for which a few new properties are deduced. For example, this 6d theory, when reduced on an $S^1$ with $\bZ_2$ twist, gives 5d $\USp(2n)$ theory with nontrivial discrete 5d theta angle.}

\end{titlepage}

\setcounter{tocdepth}{2}
\tableofcontents

\section{Introduction}
Four-dimensional $\cN=4$ super Yang-Mills theory is strongly believed to have the S-duality \cite{Montonen:1977sn,Witten:1978mh,Osborn:1979tq}, under which the theory with gauge group\footnote{In this paper we do not discuss  the global structure of the gauge group and we only keep track of the Lie algebra of the gauge group. The reader can regard any simple group used in the paper simply connected.} $G$ at coupling constant $g$ is exchanged with the theory with gauge group $G^\vee$ at coupling constant $g'\propto 1/g$.   Here $G^\vee$ is the Goddard-Nuyts-Olive dual of $G$, determined by the condition that  the Dynkin diagram of $G^\vee$ is obtained by reversing the arrows of the Dynkin diagram of $G$.\footnote{R. Langlands introduced the same operation in a rather different mathematical context, see~\cite{L}. }

This non-perturbative duality can be understood geometrically by compactifying 6d $\cN=(2,0)$ theory on $T^2$ with a suitable twist \cite{Witten:1995zh,Seiberg:1997ax,Vafa:1997mh,Ganor:1997jx}. Then the 4d coupling constant is the ratio of the two radii of $T^2$, and the S-duality is the exchange of two sides of $T^2$. In view of the recent  observations  e.g.~\cite{Douglas:2010iu,Lambert:2010iw,Bolognesi:2011rq,Bolognesi:2011nh} that most (or optimistically, all) of the BPS states of the 6d theory on $S^1$ can be found by studying the 5d maximally-supersymmetric Yang-Mills theory, it would not be completely useless to revisit the S-duality from the point of view of 5d theory on $S^1$.

We will see that maximally supersymmetric 5d theory on $S^1$ still has a kind of S-duality, which exchanges the tower of Kaluza-Klein (KK) W-bosons with the tower of instantonic monopoles. Here instantonic monopoles refer to the monopole-string configuration wrapped on $S^1$ which also has an instanton number as a configuration on the spatial $\bR^3\times S^1$.
This S-duality maps a simply-laced gauge theory on $S^1$ to itself, and a non-simply-laced gauge theory on $S^1$ to  a simply-laced gauge theory twisted by an outer automorphism around $S^1$ as follows:

\smallskip

\centerline{\begin{tabular}{rl@{ theory }ccl@{ theory twisted by }l}
1)&$\SO(2\ell+1)$&  & $\stackrel{S}{\longleftrightarrow}$ & $\SU(2\ell)$  & $\bZ_2$,  \\
2)&$\USp(2\ell)$  & with $\theta_{5d}=0$& $\stackrel{S}{\longleftrightarrow}$& $\SO(2\ell+2)$  & $\bZ_2$, \\
3)&$\USp(2\ell)$  & with $\theta_{5d}=\pi$& $\stackrel{ST}{\longleftrightarrow}$& $\SU(2\ell+1)$  & $\bZ_2$, \\
4)&$F_4$  &&$\stackrel{S}{\longleftrightarrow}$ & $E_6$  & $\bZ_2$,\\
5)&$G_2$  &&$\stackrel{S}{\longleftrightarrow}$& $\SO(8)$  & $\bZ_3$.
\end{tabular}}

\smallskip 

\noindent Here, $\theta_{5d}$ is the 5d discrete theta angle associated to $\pi_4(\USp(2\ell))=\bZ_2$; there is no theta angle for other groups, because $\pi_4$ is trivial. Note that in 4d $\cN=4$  theory, $G_2$ and $F_4$ are self-dual, and $\USp(2\ell)$ and $\SO(2\ell+1)$ are exchanged.\footnote{For subtleties in the non-simply-laced case, see \cite{Argyres:2006qr}.}  So, we can find a finer structure by considering 5d theory on $S^1$. In the following, we refer to the 5d theory on $S^1$ around which we perform the gauge transformation by an outer automorphism as a twisted 5d theory. It is to be understood that it is not related to the usual topological twisting.

Except for the case 3), this mapping follows the Langlands dual of the affine Lie algebra, obtained by reversing the direction of the arrows of the corresponding Dynkin diagram, see Fig.~\ref{untwisted} and Fig.~\ref{twisted}. Also, except for the case 3), this duality can be easily realized  in terms of 6d $\cN=(2,0)$ theory, using the following known properties:

\medskip

\begin{tabular}{@{$\bullet$ 6d theory of type\ }r@{\ on $S^1$ with\ }c@{\ twist gives 5d\ }r@{\ theory}r}
$\SU(2\ell)$&  $\bZ_2$ & $\SO(2\ell+1)$&,\\[.8ex]
$\SO(2\ell+2)$& $\bZ_2$ &  $\USp(2\ell)$&,\\[.8ex]
$E_6$ & $\bZ_2$ &  $F_4$&,\\[.8ex]
$\SO(8)$ & $\bZ_3$ &$G_2$&.
\end{tabular}

\medskip 
\noindent Note that the $\USp(2\ell)$ theory thus obtained has $\theta_{5d}=0$.
These facts have been mostly known to the experts, see e.g.~\cite{Lee:1997vp,Lee:1998vu,Kraan:1998kp,Davies:2000nw,Hanany:2001iy,Kim:2004xx,Witten:2009at,Witten:2009mh,Witten:2011zz}.  
Our discussions on these cases should then be thought of as a review. 

However, the analysis of the case 3), namely the twisted $\SU(2\ell+1)$ theory and $\USp(2\ell)$ theory with $\theta_{5d}=\pi$ turns out to be rather subtle. We approach this problem by performing the semiclassical quantization of instantonic monopole. We see that the dyons of twisted $\SU(2\ell+1)$ theory contains the spectrum of $\USp(2\ell)$ theory, and vice versa. Along the way, we give one piece of evidence of a long-standing guess that \emph{the $\USp$ theory on D4-branes on top of an $\top{4}$-plane has $\theta_{5d}=\pi$}.
Assuming the spectra thus found can be reproduced by 6d construction, we will deduce a few hitherto-unappreciated properties of 6d $\cN=(2,0)$ theory of type $\SU(2\ell+1)$ in the presence of $\bZ_2$ twist lines. For example, we find that

\medskip

\emph{6d theory of type $\SU(2\ell+1)$ on $S^1$ with $\bZ_2$ twist gives the 5d $\USp(2\ell)$ theory with $\theta_{5d}=\pi$.}

\medskip

\noindent We will leave the derivation of this and other properties from the purely 6d point of view as a possible future work.

\medskip

The rest of the paper is organized as follows. We start in Sec.~\ref{4d} by recalling how the S-duality of 4d $\cN=4$ super Yang-Mills exchanges W-bosons of gauge group $G$ with monopoles of its dual group $G^\vee$. The main tool is the embedding of the $\SU(2)$ BPS monopole into the larger gauge group.

 In Sec.~\ref{5d}, we consider 5d gauge theory on $S^1$. We study the spectrum of the KK W-bosons, and instantonic monopoles obtained by again embedding the standard 4d $\SU(2)$ BPS monopoles into the 5d theory. We will see that their charges are naturally labeled by roots of affine Lie algebras, and the S-duality involves the dual of the affine Lie algebra. We will see that an untwisted non-simply laced gauge theory is mapped to a twisted simply-laced gauge theory under the duality. 
But we find an exception that a $\bZ_2$ twisted $\SU(2\ell+1)$ theory is self-dual. 

In Sec.~\ref{partI}, we provide the geometric interpretation of the S-duality of 5d theory on $S^1$ by considering 6d $\cN=(2,0)$ theory of type $G$ on a torus, possibly with a outer-automorphism twist line. We will find that this interpretation is straightforward except for $\bZ_2$-twisted $\SU(2\ell+1)$ theory.  

To investigate this last case, in Sec.~\ref{dyons} we present a brief analysis of the semi-classical quantization of the instantonic monopoles and study the spectrum of the half-BPS dyons of 5d theory on $S^1$.  We find that $\USp(2\ell)$ theory with $\theta_{5d}=\pi$ on $S^1$ contains, in its dyon spectrum, the spectra of KK W-bosons of $\SU(2\ell+1)$ theory on $S^1$ with $\bZ_2$ outer-automorphism twist. We will also see that the spectra support the statement that $\USp$ theory on D4-branes on top of an $\top4$-plane has $\theta_{5d}=\pi$. 

We conclude the paper in Sec.~\ref{partII} by extracting the properties of 6d theory of type $\SU(2\ell+1)$ on $T^2$ with a $\bZ_2$ twist line from the dyon spectrum studied in Sec.~\ref{dyons}. The derivation of the features of the 6d theory thus extracted from string/M theory or from purely 6d analysis is left as a possible future direction. In the Appendix, we study the effect of the discrete theta angle of 5d $\USp(2\ell)$ theory to the quantization of monopole strings.

\section{S-duality of 4d theory}\label{4d}
Let us start by recalling how the S-duality exchanges the monopoles and the W-bosons. 
Consider 4d $\cN=4$ theory with gauge group $G$ at coupling constant $g$. 
Choose a Cartan subalgebra $\fh$ of the Lie algebra $\fg$ of the gauge group.
We choose the basis $E_\alpha$ of the non-Cartan part of $\fg$ such that they are simultaneous eigenstates under the action of $h\in \fh$: \begin{equation}
[h,E_\alpha]=(\alpha\cdot h) E_\alpha
\end{equation} where $\cdot$ is an invariant inner product on $\fh$. $\alpha$ is called a root of $G$. We denote the set of roots by $\Delta$.

We give a vev $\vev{\Phi}$ in the Cartan to one of the six adjoint scalars of the $\cN=4$ theory. Then the fluctuation of the component $E_\alpha$ of the gauge fields has a mass given by \begin{equation}
m_W(\alpha) = | \alpha \cdot \Phi |.
\end{equation}

Each root $\alpha$ gives a monopole solution in the following manner. Recall the standard Bogomolny-Prasad-Sommerfield monopole solution of $\SU(2)$ gauge theory, which solves the equation \begin{equation}
B_i = D_i\phi
\end{equation} where  $i=1,2,3$ is the index of spatial $\bR^3$, $B_i$ is the magnetic field constructed from the gauge field $A_i$, and $\phi$ is an adjoint scalar. 
Given the asymptotic value of $\vev{\phi}$, we have a four-parameter family of solutions; three parameters specify the center of the solution, and the last parameter is the $\U(1)$ angle. 
We normalize the coupling constant so that the mass of the solution is given by \begin{equation}
\frac{4\pi}{g^2}|\sigma^3 \cdot \vev{\phi}|.
\end{equation}  Here and in the following we take the convention that the vev of an adjoint field is measured at infinity, and gauge-rotated  to lie in a fixed Cartan algebra. 
For $\SU(2)$, the inner product on the Cartan algebra is given by the standard trace $\Tr$, so that $\sigma^3\cdot\sigma^3 =2$.

For a given root $\alpha$, $E_\alpha$, $E_{-\alpha}$ and $\alpha^\vee=2\alpha/(\alpha\cdot\alpha)$ satisfy the same commutation rules with $\sigma^+$, $\sigma^-$ and $\sigma^3$ of $\SU(2)$, thanks to the relations such as  \begin{equation}
[\alpha^\vee,E_\alpha]=2E_\alpha.
\end{equation} $\alpha^\vee$ is called a coroot. 
We will denote by $\SU(2)_\alpha$ the $\SU(2)$ subgroup defined in this manner for a root $\alpha$.
Note that   $ e^{\pi i \alpha^\vee}=-1 \in \SU(2)_\alpha$.

Using this triple, we can embed the $\SU(2)$ BPS monopole into a larger $G$, by 
decomposing the $\SU(2)$ solution  into components as \begin{equation}
\phi = \sigma^+ \phi^+ + \sigma^3 \phi^3 + \sigma^- \phi^-
\end{equation}  and then choosing the scalar field of the $G$ theory to be \begin{equation}
\Phi = \vev{\Phi}_\perp + E_\alpha \phi^+ + \alpha^\vee \phi^3 + E_{-\alpha} \phi^- 
\end{equation} and similarly for the gauge field, where we take $\sigma^3\cdot \vev{\phi}=\alpha\cdot\vev{\Phi}$ and $\vev{\Phi}_\perp$ is the projection of $\vev{\Phi}$ orthogonal to $\alpha$. 
The classical mass of this configuration is then given by  \begin{equation}
m_M(\alpha)=\frac{4\pi}{g^2} \frac{\alpha^\vee\cdot\alpha^\vee}{\sigma^3\cdot \sigma^3}|\sigma^3\cdot \phi|=   \frac{4\pi}{g^2} |\alpha^\vee\cdot\Phi|.
\end{equation}

Both the W-bosons and the monopoles thus constructed preserve half of the supersymmetry, and therefore the mass does not receive quantum corrections. In $\cN=4$ theory, the quantization of the fermionic zero modes around the monopole makes it into an $\cN=4$ massive vector multiplet, just as a W-boson is \cite{Osborn:1979tq}.

Let $G^\vee$ be a group whose Dynkin diagram is obtained by reversing the arrow of the Dynkin diagram of $G$.  Then its set of roots $\Delta(G^\vee)$ is the set of coroots $\Delta^\vee(G)$ of $G$ up to an overall scaling: \begin{equation}
\Delta^\vee(G)= n_G\Delta(G^\vee).
\end{equation}  

Then the spectrum of the W-bosons and the monopoles of the theory with gauge group $G$ at the coupling constant $g$ and that of the theory with gauge group $G^\vee$  at the coupling constant $g'$ are the same if \begin{equation}
\frac{4\pi}{g'{}^2} = \frac1{n_G}\frac{g^2}{4\pi}
\end{equation} under the identification \begin{align*}
\text{W-bosons of $G$} &\leftrightarrow \text{monopoles of $G^\vee$}, \\
\text{monopoles of $G$} &\leftrightarrow \text{W-bosons of $G^\vee$}.
\end{align*} 
The fact presented above is only a very superficial part of the S-duality of $\cN=4$ theory.  For more, see e.g.~\cite{Sen:1994yi,Vafa:1994tf,Girardello:1995gf} from the previous centruy, or e.g.~\cite{Gadde:2009kb,Spiridonov:2010qv}.

\section{S-duality of 5d theory on $S^1$}\label{5d}
\subsection{Kaluza-Klein W-bosons and instantonic monopoles}
Now let us move on to the analysis of the 5d maximally-supersymmetric Yang-Mills theory.
The theory is specified by the choice of the gauge group $G$ and the coupling constant $g$ which now has a dimension. When $G=\USp(2\ell)$ we also need to specify the discrete theta angle corresponding to $\pi_4(\USp(2\ell))=\bZ_2$; for simplicity we take it to be zero in this section. We will come back to the nonzero case in Sec.~\ref{dyons}.

The theory is not consistent per se, at least in the standard interpretation. We suppose we have chosen some stringy or M-theoretic ultraviolet completion of the theory. 
Let us put this theory on $S^1$ with coordinate $x^5$, of circumference $R_5$. The 5d theory has five adjoint scalars. Together with the Wilson line $\int A_5 dx^5$ around $S^1$, they comprise the six adjoint scalars of the 4d theory in the low energy limit.  For simplicity let us turn on the vev $\Phi$ to one scalar in the five adjoint scalars of the 5d theory, keeping the Wilson line trivial.

Now the W-bosons come in the KK tower, so the charge of a W-boson is labeled by a pair \begin{equation}
\aff\alpha=(k,\alpha)
\end{equation} where $k$ is an integer specifying the KK momentum. Its mass is given by \begin{equation}
m_W(\aff\alpha)=\sqrt{(\frac{2\pi k}{R_5})^2+|\alpha\cdot\Phi|^2} = 
|\frac{2\pi i k}{R_5} + \alpha\cdot \Phi|.
\end{equation} 
Therefore, defining $\aff\Phi=(2\pi i/R_5,\Phi)$, we have
\begin{equation}
m_W(\aff\alpha)=|\aff\alpha \cdot \aff\Phi|.
\end{equation}

Can there be a kind of S-duality in this theory? If so, monopoles also need to form a tower, and indeed they do \cite{Lee:1997vp,Lee:1998vu,Kraan:1998kp,Davies:2000nw,Hanany:2001iy,Kim:2004xx}. 
The trick is first to embed the basic Bogomolny-Prasad-Sommerfield $\SU(2)$ monopole solution to the theory in five-dimension using both $\Phi$ and $A_5$, so that both are independent of $x^5$:
\begin{alignat}{3}
\Phi&=\vev{\Phi}_\perp &+& \cos\theta &(E_\alpha\phi^++\alpha^\vee \phi^3+E_{-\alpha}\phi^-), \\
A_5&= &+& \sin\theta &(E_\alpha\phi^++\alpha^\vee \phi^3+E_{-\alpha}\phi^-).
\end{alignat} where we take the standard $\SU(2)$  monopole solution $\phi$ such that   
$u=\sigma^3\cdot\vev\phi$ satisfies \begin{equation}
u \cos \theta = \alpha\cdot\vev{\Phi}, \qquad
u \sin\theta = \alpha\cdot \vev{A_5}=\frac{2\pi k}{R_5}
\end{equation} for a root $\alpha$. 
This gives a solution of the Bogomolny equation locally on the spacetime, because $\Phi$ and $A_5$ behave as 4d scalars as long as they are independent of $x^5$.

The holonomy around $S^1$ at the asymptotic infinity is $(-1)^k \in \SU(2)_\alpha \subset G$. Therefore, this solution for odd $k$ is not a state in the same vacuum for a simply-connected $G$.
To cure this problem, we perform a large gauge transformation by \begin{equation}
g(x_5)=e^{\pi k x_5 \alpha^\vee /R_5}
\end{equation} so that $g(0)=1$, $g(R_5)=(-1)^k$.  We now have
\begin{align}
\Phi&=\vev{\Phi}_\perp + \cos\theta (E_\alpha(x_5)\phi^++\alpha^\vee \phi^3+E_{-\alpha}(x_5)\phi^-), \label{config1}\\
A_5&= \sin\theta (E_\alpha(x_5)\phi^+ +\alpha^\vee (\phi^3 -u/2) +E_{-\alpha}(x_5)\phi^-)
\label{config2}
\end{align} where 
\begin{equation}
E_{\pm\alpha}(x_5)= g(x_5) E_{\pm\alpha}  g(x_5)^{-1}.
\end{equation}
Then for every integral $k$ this solution determines a state in the proper vacuum we are interested.

The mass of the solution is then given by \begin{equation}
m_M(\aff\alpha)=\frac{4\pi R_5} {g^2} \frac{\alpha^\vee\cdot\alpha^\vee}{\sigma^3\cdot\sigma^3} |u|
=\frac{4\pi R_5}{g^2} |\aff\alpha^\vee \cdot \aff\Phi|
\end{equation} where $\aff\alpha$ and $\aff\Phi$ were as defined above, and \begin{equation}
\aff\alpha^\vee=\frac{2}{\alpha\cdot\alpha}\aff\alpha 
=(\frac{2k}{\alpha\cdot\alpha}, \alpha^\vee).
\end{equation} 
Note that $2k/(\alpha\cdot\alpha)$ in this monopole solution measures the instanton number on the spatial slice $\bR^3\times S^1$: \begin{equation}
\#\text{instanton}\propto\int_{\bR^3\times S^1} \Tr F\wedge F
\propto\int d\theta d\varphi B_{\theta\varphi}\int dr dx^5 \partial_r A_5 \propto \frac{2k}{\alpha\cdot\alpha}.
\end{equation}
Therefore the tower is of the instantonic monopole strings wrapped around $S^1$.

Summarizing, the mass spectrum of two towers, one of KK W-bosons and another of instantonic monopole strings, is given by \begin{equation}
m_W(\aff\alpha)=|\aff\alpha\cdot\aff\Phi|,\qquad
m_M(\aff\alpha)=\frac{4\pi R_5}{g^2}|\aff\alpha^\vee \cdot\aff\Phi|.
\end{equation}
When $G$ is simply-laced, $\aff\alpha^\vee=\aff\alpha$. Therefore,
 the mass spectra of the theory with simply-laced gauge group $G$ on $S^1$
 at coupling constant $g$ and the radius $R_5$ and
 at coupling constant $g'$ and the radius $R_5'$  
are identical  after the exchange of the KK W-bosons and the instantonic monopoles
 when \begin{equation}
\frac{4\pi R_5'}{g'{}^2} = \frac{g^2 }{4\pi R_5}.
\end{equation}
But the situation for non-simply-laced $G$ is slightly more complicated. 
The situation is best understood by working out an example.

\subsection{An example: $\USp(2\ell)$ $\leftrightarrow$ twisted $\SO(2\ell+2)$ }
Let us take $G=\USp(2\ell)$. The root space is $\ell$ dimensional. Introduce an orthogonal basis $e_i$, $i=1,\ldots,\ell$. Then the set of roots is given by $\pm 2e_i$ and $\pm e_i \pm e_j$ for $i\ne j$.
Therefore the set of $\aff\alpha$ is comprised of \begin{equation}
(k,\pm 2e_i),\quad (k,\pm e_i \pm e_j).
\end{equation} Then the set of $\aff\alpha^\vee$ is made of \begin{equation}
(k+\frac12,\pm e_i),\quad (k,\pm e_i),\quad (k,\pm e_i \pm e_j) \label{foo}
\end{equation}  where $k$ is an integer. Clearly, this set cannot be the set of charges of KK W-bosons of a 5d theory on $S^1$ without Wilson line.

There is, however, a construction for which the charges of KK W-bosons are the ones listed in \eqref{foo}.  Take a 5d super Yang-Mills theory with gauge group $\SO(2\ell+2)$, and put it on $S^1$ with the boundary condition \begin{equation}
\phi(x^5+ R_5) = P \phi(x^5) P^{-1}
\end{equation} where $P$ is the parity transformation of $\mathrm{O}(2\ell+2)$.
The adjoint representation decomposes under $P$ as \begin{equation}
\mathfrak{so}(2\ell+2) =\mathfrak{so}(2\ell+1) \oplus \underline{2\ell+1}
\end{equation} where $P$ acts by a multiplication by $-1$ on the vector representation $\underline{2\ell+1}$ of the $\SO(2\ell+1)$ subgroup which is invariant under $P$.
The roots of $\SO(2\ell+1)$ are $\pm e_i$ and $\pm e_i\pm e_j$, giving the KK modes labeled by $(k,\pm e_i)$ and $(k,\pm e_i\pm e_j)$ in \eqref{foo}. 
The weights of the vector representation $\underline{2\ell+1}$ are $\pm e_i$. They are odd under $P$, therefore carry a half-integral KK momentum, thus giving modes with charges $(k+\frac12,\pm e_i)$ in \eqref{foo}.

As for the instantonic monopoles of a twisted theory, one finds that if $(k,\alpha)$ is a charge for a KK W-boson where $k$ is in general non-integral, the monopole configuration constructed in \eqref{config1}, \eqref{config2} still satisfies the appropriate twisted boundary condition, with the magnetic charge $(k,\alpha)^\vee$. 
In particular, the spectrum of the instantonic monopoles of the twisted $\SO(2\ell+2)$ theory matches with that of the KK W-bosons of the original $\USp(2\ell)$ theory.

This observation can be generalized: the S-dual of 5d super Yang-Mills with non-simply-laced gauge group $G$ on $S^1$ is given by 5d super Yang-Mills with simply laced gauge group $\tilde G$ with twist around $S^1$ by an automorphism. To understand this statement fully, it is instructive to recall the structure of twisted affine Lie algebras.

\subsection{Twisted affine Lie algebras}

\begin{figure}
\[
\begin{array}{r@{\quad}l@{\qquad}l@{\quad}l}
\SU(2)=A_1&\node{}{\alpha_1}&
A_1^{(1)} & \node{}{1} \Leftrightarrow \node{}{1} \\
\SU(\ell+1)=A_\ell & \node{}{\alpha_1} - \node{}{\alpha_2} - \cdots - \node{}{\alpha_{\ell-1}}-\node{}{\alpha_\ell} &
A_\ell^{(1)}  & 
\begin{array}{c}
\raisebox{-12pt}{\rotatebox{30}{$-\!\!-\!\!-$}}\node{}{1}\raisebox{0pt}{\rotatebox{-30}{$-\!\!-\!\!-$}} \\[-7pt]
\node{}{1}-\node{}{1}-\cdots-\node{}{1} 
\end{array}\\
\SO(2\ell+1)=B_\ell & \node{}{\alpha_1} - \node{}{\alpha_2} - \cdots - \node{}{\alpha_{\ell-1}} \Rightarrow\node{}{\alpha_\ell} \tikz[na]\node(B){};&
B_\ell^{(1)}  & \node{}{1}-\node{\ver{}{1}}{2}-\node{}{2}-\cdots-\node{}{2}\Rightarrow\node{}{2} \ \tikz[na]\node(B1){};\\
\USp(2\ell)=C_\ell & \node{}{\alpha_1} - \node{}{\alpha_2} - \cdots - \node{}{\alpha_{\ell-1}} \Leftarrow\node{}{\alpha_\ell} \tikz[na]\node(C){};&
C_\ell^{(1)} & \node{}{1}\Rightarrow \node{}{2}-\cdots-\node{}{2}\Leftarrow\node{}{1} \ \tikz[na]\node(C1){};\\
\SO(2\ell)=D_\ell& \node{}{\alpha_1} - \node{}{\alpha_2} - \cdots - \node{\ver{}{\alpha_{\ell-1}}}{\alpha_{\ell-2}} -\node{}{\alpha_{\ell-1}} &
D_\ell^{(1)}  & \node{}{1}-\node{\ver{}{1}}{2}-\node{}{2}-\cdots-\node{\ver{}{1}}{2}-\node{}{1} \\
E_6& \node{}{\alpha_1} - \node{}{\alpha_2} - \node{\ver{}{\alpha_{6}}}{\alpha_{3}} -\node{}{\alpha_{4}} -\node{}{\alpha_5} &
E_6^{(1)}  & \node{}{1}-\node{}{2}-\node{\overset{\ver{}{1}}{\ver{}{2}}}{3}-\node{}{2}-\node{}{1} \\
E_7& \node{}{\alpha_1} - \node{}{\alpha_2} - \node{\ver{}{\alpha_{7}}}{\alpha_{3}} -\node{}{\alpha_{4}} -\node{}{\alpha_5} -\node{}{\alpha_6} &
E_7^{(1)} & \node{}{1}-\node{}{2}-\node{}{3}-\node{\ver{}{2}}{4}-\node{}{3}-\node{}{2}-\node{}{1}\\
E_8& \node{}{\alpha_1} - \node{}{\alpha_2} - \node{}{\alpha_{3}} -\node{}{\alpha_{4}} -\node{\ver{}{\alpha_{8}}}{\alpha_5} -\node{}{\alpha_6}-\node{}{\alpha_7} &
E_8^{(1)} & \node{}{1}-\node{}{2}-\node{}{3}-\node{}{4}-\node{}{5}-\node{\ver{}{3}}{6}-\node{}{4}-\node{}{2} \\
F_4 & \node{}{\alpha_1} - \node{}{\alpha_2} \Rightarrow \node{}{\alpha_3} -\node{}{\alpha_4}&
F_4^{(1)} & \node{}{1}-\node{}{2}-\node{}{3}\Rightarrow\node{}{4}-\node{}{2} \quad\tikz[na]\node(F41){};\\
G_2 & \node{}{\alpha_1} \Rrightarrow \node{}{\alpha_2} &
G_2^{(1)} & \node{}{1}-\node{}{2}\Rrightarrow\node{}{3} \quad\tikz[na]\node(G21){};
\end{array}
\]
\caption{Finite and untwisted affine Dynkin diagrams. The labels of the untwisted diagram are the same with the corresponding finite diagram. We call the extending node $\alpha_0$. \label{untwisted}}
%\end{figure}

%\begin{figure}

\begin{align*}
A_2^{(2)} &&& \node{2}{\alpha_0} \Llleftarrow \node{1}{\alpha_1} \\
A_{2\ell}^{(2)}  &&& \node{2}{\alpha_0}\Leftarrow \node{2}{\alpha_1}-\cdots-\node{2}{\alpha_{\ell-1}}\Leftarrow\node{1}{\alpha_{\ell}} \\
A_{2\ell-1}^{(2)}  &&& \node{1}{\alpha_1}-\node{\ver{1}{\alpha_0}}{\alpha_2}\!\!{}^2-\node{2}{\alpha_3}-\cdots-\node{2}{\alpha_{\ell-1}}\Leftarrow\node{1}{\alpha_\ell} \ \tikz[na]\node(Aodd2){}; \\
D_{\ell+1}^{(2)}  &&& \node{1}{\alpha_0}\Leftarrow \node{1}{\alpha_1}-\cdots-\node{1}{\alpha_{\ell-1}}\Rightarrow\node{1}{\alpha_\ell} \quad\tikz[na]\node(D2){};\\
E_6^{(2)} &&& \node{1}{\alpha_0}-\node{2}{\alpha_1}-\node{3}{\alpha_2}\Leftarrow\node{2}{\alpha_3}-\node{1}{\alpha_4} \quad\tikz[na]\node(E62){};\\
D_4^{(3)} &&& \node{1}{\alpha_0}-\node{2}{\alpha_1}\Lleftarrow\node{1}{\alpha_3} \quad\tikz[na]\node(D43){};
\end{align*}

\begin{tikzpicture}[overlay]
        \path[<->,style=mine] (B1) edge [color=red,out=-60,in=60] (Aodd2);
        \path[<->,style=mine] (C1) edge [color=green,out=-60,in=30] (D2);
        \path[<->,style=mine] (F41) edge [color=blue,out=-30,in=15] (E62);
        \path[<->,style=mine] (G21) edge [color=orange,out=-30,in=0] (D43);
\end{tikzpicture}

\caption{Twisted affine Dynkin diagrams. Action of the Langlands duality is superimposed on top. Note that $A^{(2)}_\text{even}$ is self-dual up to the reversal of the order of the vertices.\label{twisted}}
\end{figure}

Consider currents $j(s,t)$ of group $G$ on a 2d cylinder, with periodicity $s\sim s+2\pi$.
When the boundary condition on $j$ around the $s$ direction is trivial, the expansion of the currents into modes gives rise to operators $J^a_k$ for integral $k$ and $a=1,\ldots,\dim G$, which form the \emph{untwisted} affine Lie algebra $G^{(1)}$.  Its roots have the form $(k,\alpha)$ where $\alpha$ is a root of $G$, and its Dynkin diagram is the extended Dynkin diagram of $G$, see Fig.~\ref{untwisted}.   
Note that the charges $(k,\alpha)$ of the KK W-bosons of the 5d theory with gauge group $G$ on $S^1$ are exactly the affine roots $\aff\alpha$ of $G^{(1)}$. 
We used $\aff\alpha^\vee=2\aff\alpha/(\alpha\cdot\alpha)$, which are the affine coroots.
This means that it is natural to use the metric on the vectors $\aff\alpha=(k,\alpha)$ such that the direction measured by $k$ is null, which is also in accord with the usage of affine Lie algebras. 
As the inner product on $\aff\alpha$ is degenerate, one cannot identify the root space with its dual, so $\aff\Phi$ needs to be regarded as an element of the dual of the root space.

We can impose a more general boundary condition \begin{equation}
j(s+2\pi) = \sigma(j(s))\label{twist}
\end{equation} where $\sigma$ is an element of $\Aut(G)$, i.e.~an automorphism of $G$.
Then the modes of the currents can have non-integral mode numbers depending on the eigenvalues of $\sigma$. 

\begin{figure}
\def\Node#1#2{\overset{#1}{\tikz[na]\node(#2){$\displaystyle\circ$};}}
\def\Ver#1#2{\overset{{\llap{$\scriptstyle#1$}\tikz[na]\node(#2){$\displaystyle\circ$};}}{\scriptstyle\vert}}
\[
\begin{array}{lclc}
A_\text{odd} &  \Node{}{a}-\Node{}{}-\Node{}{b} &
A_\text{even} & \Node{}{c}-\Node{}{d}-\Node{}{e}-\Node{}{f} \\
D_n & \Node{}{}-\Node{}{}-\Node{\Ver{}{g}}{}-\Node{}{h}  &
E_6 & \Node{}{i}-\Node{}{j}-\Node{\Ver{}{}}{}-\Node{}{k}-\Node{}{l} \\
D_4 &  \Node{}{m}-\Node{\Ver{}{n}}{}-\Node{}{o}
\end{array}
\]

\begin{tikzpicture}[overlay,thin,arrows={angle 90}-{angle 90}]
        \path[] (a) edge[out=60,in=120]  (b);
        \path[] (c) edge[out=60,in=120]  (f);
        \path[] (d) edge[color=red,out=60,in=120]  (e);
        \path[] (g) edge[out=0,in=90]  (h);
        \path[] (i) edge[out=-60,in=-120]  (l);
        \path[] (j) edge[out=-60,in=-120]  (k);
        \path[-{angle 90}] (m) edge[out=-60,in=-120]  (o);
        \path[-{angle 90}] (o) edge [out=90,in=0] (n);
        \path[-{angle 90}] (n) edge [out=180,in=90] (m);
\end{tikzpicture}

\caption{Graph automorphisms of finite simply-laced Dynkin diagrams, or equivalently the outer automorphisms of the corresponding Lie algebras. See the arrow emphasized in red: only for $A_\text{even}=\SU(2\ell+1)$ there are two nodes which are  connected both by a line in the Dynkin diagram and an arrow representing the action of the automorphism.   \label{outer}}
\end{figure}
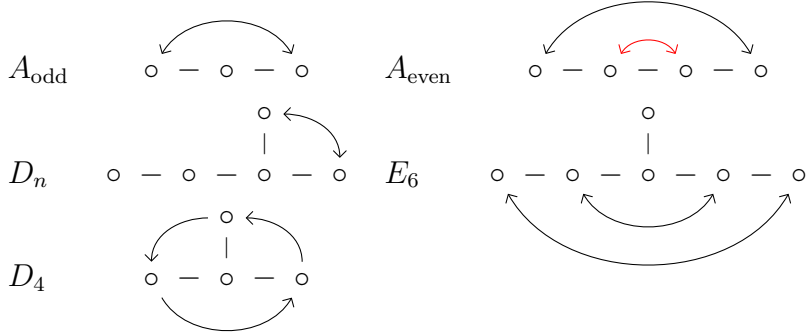

For each element $g\in G$ we have an inner automorphism $\sigma_g\in\Inn(G)$ given by $\sigma_g(x)=gxg^{-1} \in \Inn(G)$.
The coset $\Out(G)=\Aut(G)/\Inn(G)$, where we identify $\sigma$ and $\sigma \sigma_g$, is the outer automorphism group. Let us denote by $[\sigma]$ the projection of $\sigma$ to the outer automorphism group. 
It is a classic theorem that $\Out(G)$ can be identified with the graph automorphism of the Dynkin diagram of $G$, see Fig.~\ref{outer}.

When $[\sigma]$ is trivial, the Lie algebra formed by the modes is still isomorphic to the original untwisted affine Lie algebra $G^{(1)}$. When $[\sigma]$ is nontrivial and is of order $r=2,3$, the resulting Lie algebra is the \emph{twisted} Lie algebra $G^{(r)}$, whose Dynkin diagrams are given in Fig.~\ref{twisted}.  For example, the vectors \eqref{foo} labeling the spectrum of KK W-bosons of $\SO(2\ell+2)$ theory with a parity twist are the roots of $D_{\ell+1}^{(2)}$. 

For a given affine Lie algebra $G^{(r)}$, the set of its coroots $\aff\alpha^\vee$ is the set of roots of an affine Lie algebra $\tilde G^{(\tilde r)}$, which can be found by reversing the arrows of the Dynkin diagram.  This operation is called the Langlands dual of the affine Lie algebra, and is denoted by $\tilde G^{(\tilde r)}=(G^{(r)}) ^\vee$. 
Note that $\tilde G$ is not always $G^\vee$. For example, $F_4^\vee=F_4$ but $(F_4^{(1)})^\vee=E_6^{(2)}$.

We are interested in a finer classification which keeps track of the mode numbers of the currents,   not just the Lie algebra formed by the modes. 
 For example, all inner automorphism $\sigma_g$ for $g\ne 1$ is trivial as an outer automorphism, but a nontrivial $\sigma_g$ still introduces non-integral mode numbers.
Instead, we regard two automorphism $\sigma$ and $\sigma'$ as conjugate when there is another automorphism $\tau$ such that $\sigma'=\tau\sigma\tau^{-1}$. The mode expansions are the same when the automorphisms $\sigma$ appearing in \eqref{twist} are conjugate.
Conjugacy classes of $\sigma$ of finite order can be described using Kac's theorem from \S8.6 of Kac's textbook \cite{Kac}: 
\begin{itemize}
\item Given an assignment of non-negative integers $(s_0,\ldots,s_\ell)$, at least one of which is nonzero, to the nodes of the Dynkin diagram of type $G^{(r)}$ such that their greatest common divisor is one, there is a standard automorphism $\sigma_r(s_0,\ldots,s_\ell)$ of $G$.\footnote{This was denoted by $\sigma_{s_0,\ldots,s_\ell;r}$ in the textbook \cite{Kac}.} Its order as an outer automorphism is $r$, and its order as an automorphism is $s=r\sum_i s_i n_i$ where $n_i$ are the integers assigned to the nodes in Fig.~\ref{untwisted} and Fig.~\ref{twisted}.   
\item Any nontrivial finite order automorphism $\sigma$ of $G$ is conjugate to a standard outer automorphism.
\item Two standard automorphisms $\sigma_r(s_0,\ldots,s_\ell)$ and $\sigma_r(s_0',\ldots,s_\ell')$ are conjugate if and only if the numbers $s_i$ and $s_i'$ are mapped to each other by a symmetry of the Dynkin diagram of $G^{(r)}$.
\item The Dynkin diagram of the subgroup of $G$ invariant under $\sigma(s_0,\ldots,s_\ell)$ is given by the sub-Dynkin diagram of $G^{(r)}$ consisting of the nodes for which $s_i=0$ and edges connecting them. 
\end{itemize}

Let us illustrate the theorem by a few examples. For brevity, we denote by $\sigma_r(k)$ the standard automorphism $\sigma_r(\{s_i\})$ where $s_k=1$ and $s_i=0$ otherwise. 
Then for any $G$, the identity is $\sigma_1(0)$. 
Next, take $G=\SO(2\ell+2)$.
The standard automorphism $\sigma_2(k)$ is given by the parity operation
\begin{equation}
x \mapsto P_k x P_k^{-1} \qquad\text{where}\qquad
P_k=\diag(\underbrace{-1,\ldots,-1}_{2k+1},+1,\ldots,+1). \label{aaa}
\end{equation}
The order is always two, the subgroup invariant under it is $\SO(2k+1)\times \SO(2(\ell-k)+1)$, and $\sigma_2(k)$ is the same with $\sigma_2({\ell-k})$.

The Langlands dual of $\SO(2\ell+2)^{(2)}$ is $\USp(2\ell)^{(1)}$. 
The standard automorphism $\sigma_1(k)$ is given by the parity operation
\begin{equation}
x \mapsto Q_k x Q_k^{-1} \qquad\text{where}\qquad
Q_k=\diag(\underbrace{-1,\ldots,-1}_{2k},+1,\ldots,+1).\label{bbb}
\end{equation}    
The order is 2 unless $k=0$ or $k=\ell$, for which case the order is one.  The subgroup invariant under it is $\USp(2k)\times \USp(2(\ell-k))$. 

As a final example, consider the outer automorphisms of $\SU(2\ell+1)$. The standard automorphism $\sigma_2(k)$ is given by \begin{equation}
x\mapsto R_k (-x^t) R_k^{-1} \quad\text{where}\quad
R_k=\diag(\underbrace{+1,\ldots,+1}_{2k+1})\oplus \begin{pmatrix}
0&1\\
-1&0
\end{pmatrix} \oplus\cdots\oplus \begin{pmatrix}
0&1\\
-1&0
\end{pmatrix}. \label{bar}
\end{equation}
The order of $\sigma_2(k)$ is four unless $k=\ell$, for which the order is two. 
The subgroup invariant under it is $\SO(2k+1)\times \USp(2(\ell-k))$.

\subsection{General statement}\label{affineLanglands}
Let us consider 5d theory with gauge group $G$ with coupling constant $g$, put on a circle of radius $R_5$ with the boundary condition $\phi(x_5+ R_5)=\sigma(\phi(x_5))$ where $\sigma$ is an automorphism whose order as an outer automorphism is $r$. For simplicity we assume $\sigma$ is of finite order.  For brevity denote this setup by $(G,\sigma_r,g,R_5)$.
The charges of the KK W-bosons have the form \begin{equation}
\aff\alpha=(n+\nu,\alpha)
\end{equation} where $\alpha$ is in the part of the Cartan of $G$ which is invariant under $\sigma$, $\nu$ is the shift in the KK momentum given by $\alpha$, and $n$ is an integer. 
The mass of the KK W-boson is given by \begin{equation}
m_W(\aff\alpha)=|\aff\alpha\cdot\aff\Phi|.
\end{equation}  Similarly, the mass of instantonic monopole labeled by the same $\aff\alpha$ is given by \begin{equation}
m_M(\aff\alpha)=\frac{4\pi R_5}{g^2} |\aff\alpha^\vee\cdot\aff\Phi|.
\end{equation}

Then the following statement holds: the mass spectra of the setup $(G,\sigma_r,g,R_5)$
and $(\tilde G,\tilde \sigma_{\tilde r},\tilde g,\tilde R_5)$ are the same after the exchange of the KK W-bosons and the instantonic monopoles when the following three conditions hold:
\begin{itemize}
\item $G^{(r)}$ and $\tilde G^{(\tilde r)}$ are Langlands dual, i.e.~obtained by reversing the arrows, 
\item $\sigma_r$ and $\tilde\sigma_{\tilde r}$ are specified by the same set of integers, i.e.~they are conjugate to the standard automorphisms $\sigma_{r}=\sigma_r(s_0,\ldots,s_\ell)$ and $\tilde\sigma_{\tilde r}=\sigma_{\tilde r}(s_0,\ldots,s_\ell)$  with the same assignment of integers $(s_i)$ after the reversal of the arrows,
\item The radii and the coupling constants are related by $\displaystyle \frac{4\pi R_5}{g^2} = \frac{1}{n_{G^{(r)}}} \frac{\tilde g^2}{4\pi \tilde R_5}$ where $n_{G^{(r)}}$ is determined so that $\Delta^\vee(\tilde G^{(\tilde r)}) = n_{G^{(r)}} \Delta(G^{(r)})$.
\end{itemize}

Instead of giving a straightforward proof ridden with indices, let us illustrate the statement with a few examples.
\begin{itemize}
\item When $G$ is simply-laced, the Langlands dual of $G^{(1)}$ is itself, and the automorphism $\sigma$ does not change either. Therefore, the S-dual of 5d theory with simply-laced gauge group $G$ with the twist $\sigma$ is the same theory itself, with $R_5$ and $g^2$ interchanged. 
\item When $G$ is non-simply-laced, the Langlands dual of $G^{(1)}$ is $\tilde G^{(r)}$ for a simply-laced $\tilde G$ where $r$ is the ratio of the squared length of the long and the shoft roots. Consider 5d theory with gauge group $G$ on $S^1$ without a twist, or equivalently with a trivial twist $\sigma_1(0)$. Then the S-dual is 5d theory with gauge group $\tilde G$ on $S^1$ with an outer automorphism twist $\sigma_r(0)$, with $R_5$ and $g^2$ reversed. So, the S-dual of untwisted non-simply-laced theory is a  simply-laced theory twisted by an outer automorphism.
\item More specifically, consider 5d $\SO(2\ell+2)$ theory on $S^1$ with the outer-automorphism twist $\sigma_2(k)$ given in \eqref{aaa}. The 4d gauge group is $\SO(2k+1)\times \SO(2(\ell-k)+1)$.
Its S-dual is the 5d $\USp(2\ell)$ theory on $S^1$ with the inner-automorphism twist $\sigma_1(k)$ given in \eqref{bbb}. The 4d gauge group is $\USp(2k)\times \USp(2(\ell-k))$.
\item However, the S-dual of twisted simply-laced theory is not always an untwisted non-simply-laced theory. Consider 5d theory with gauge group $G=\SU(2\ell+1)$ on $S^1$ with a twist $\sigma_2(k)$, explicitly given by \eqref{bar}. The 4d gauge group is $\SO(2k+1)\times \USp(2(\ell-k))$ as described there.
The Langlands duality of $\SU(2\ell+1)^{(2)}$ is itself, but the node $\alpha_k$ is mapped to $\alpha_{\ell-k}$. Therefore, the S-dual is 5d theory with gauge group $G=\SU(2\ell+1)$ on $S^1$ with a twist by $\sigma_2({\ell-k})$,
whose 4d gauge group is $\SO(2(\ell-k)+1)\times \USp(2k)$, as it should be.
\end{itemize}

In a certain sense, the compactification of 5d theory on $S^1$ is a 4d theory whose gauge group is the loop group of $G$, i.e.~a group formed by a map from $S^1$ to $G$, possibly with a boundary condition given by an automorphism $\sigma$. Therefore, the S-duality should involve the Langlands dual of the loop group, giving a heuristic interpretation of the statements we found in this section.

\section{Interpretation via 6d $\cN=(2,0)$ theory, part I}\label{partI}
In this section, we show that the S-duality of 5d theory on $S^1$ we have seen in the previous sections has a simple geometric interpretation in terms of 6d $\cN=(2,0)$ theory, \emph{except for $\SU(2\ell+1)$ theory with $\bZ_2$ twist}.

The type of 6d theory is specified by a simply-laced Dynkin diagram of type $G=A_n$, $D_n$ or $E_n$.
One way to realize these interacting 6d theories is to take  a low-energy decoupling limit of Type IIB string on $\bC^2/\Gamma(G)$, where $\Gamma(G)$ is a discrete subgroup of $\SU(2)$  of type $G$ \cite{Johnson:1996py}. 
It has a Coulomb branch direction corresponding to resolving $\bC^2/\Gamma(G)$. 
In the resolved $\bC^2/\Gamma(G)$, there are two-cycles $D_i$ ($i=1,\ldots,\rank G)$  corresponding to the nodes $\alpha_i$ of  the Dynkin diagram of type $G$. 
Then the 6d theory has a string excitation of type $\alpha_i$, 
given by wrapping a D3-brane wrapped on $D_i$. This couples to a self-dual two-form potential $B^i_{(2)}$ which arise from components of the self-dual four-form of the Type IIB theory, $C_{(4)}= B^i_{(2)}\wedge \omega_{i,(2)}$ where $\omega_{i,(2)}$ is a two-form which is dual to the cycles $D_i$.

Its compactification on $S^1$ with radius $R_6$, when probed at a suitable energy scale, is a 5d maximally supersymmetric Yang-Mills with gauge group $G$ with coupling constant $4\pi/g^2=1/R_6$. This means that the KK momentum along the sixth direction is the instanton charge of the 5d theory.
A string of type $\alpha$ wrapped around the circle is a W-boson of charge $\alpha$. 
A string of type $\alpha$ not wrapped around the circle is a monopole string of charge $\alpha$.

Now let us make a further compactification on $S^1$ with radius $R_5$. This is 5d theory with gauge group $G$ on $S^1$ without any twist, and is 6d theory of type $G$ on $T^2$. 
Then the KK W-bosons of charge $(k,\alpha)$ are strings of charge $\alpha$ wrapped around the sixth direction, which has $k$ units of  KK momentum along the fifth direction.
The instantonic monopoles of charge $(k,\alpha)$ are strings of charge $\alpha$ wrapped around the fifth direction, which has $k$ units of  KK momentum along the sixth direction, or equivalently $k$ units of the instanton charge in the 5d viewpoint.
The twist by an inner automorphism $\sigma_1(s_0,\ldots,s_\ell)$ can be easily incorporated; we just demand $\int_{D_i\times T^2} C_{(4)}=2\pi s_i/s$ where $s=\sum n_i s_i$. This affects the quantization of the KK momentum in the expected way.

To incorporate the twist by an outer automorphism, we use the fact that when the finite Dynkin diagram of type $G$ has a $\bZ_r$ symmetry $\sigma$ mapping node $\alpha_i$ to $\alpha_{\sigma(i)}$, the corresponding orbifold $\bC^2/\Gamma(G)$ also has a $\bZ_r$ isometry which preserves the hyperk\"ahler structure, mapping the corresponding cycle $D_i$ to $D_{\sigma(i)}$, \emph{unless $G=\SU(2\ell+1)$,} c.f. the table of the symmetry of Dynkin diagrams in Fig.~\ref{outer}.  
 This isometry has been used many times in string theory, e.g.~\cite{Aspinwall:1996nk,Vafa:1997mh,Witten:1997kz,Vafa:1998vs,Aspinwall:1998xj}.
For the $\bZ_2$ action on the Dynkin diagram of $G=\SU(2\ell+1)$, the corresponding hyperk\"ahler isometry has order 4. We will come back to this in Sec.~\ref{partII}.

\begin{figure}
\[
\includegraphics[scale=.4]{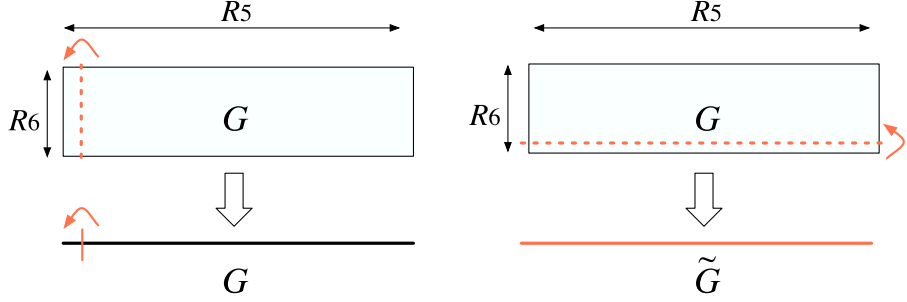}
\]
\caption{6d theory of type $G$ put on $T^2$ with twists on  different cycles, and its interpretation as 5d theory.  The orange dotted line is the place across which the $\bZ_r$ twist is performed. \label{aho}}
\end{figure}

The geometric symmetry of $\bC^2/\Gamma(G)$ guarantees that the 6d theory of type $G$ has a $\bZ_r$ symmetry \emph{unless $G=\SU(2\ell+1)$}.
Then, when we compactify the 6d theory on $T^2$, we can arrange so that the theory is twisted by this $\bZ_r$ symmetry when one goes around the fifth direction or the sixth direction, see Fig.~\ref{aho}. 
When one has a twist when one goes around the fifth direction and $R_5\gg R_6$, locally we can first go to the 5d description where we have $G$ gauge symmetry. One goes around the fifth direction, a W-boson of charge $\alpha_i$ is turned into a W-boson of charge $\alpha_{\sigma(i)}$. Furthermore, this is a $\bZ_r$ operation. This uniquely fixes that the twist of the 5d theory is via the automorphism $\sigma_r(0)$.
Its S-dual description is obtained by exchanging the fifth and the sixth directions. When $R_5\gg R_6$, this should be a 5d theory on $S^1$ without a twist. From the analysis in the last section, we know that the 5d gauge group $\tilde G$ is such that $\tilde G^{(1)}=(G^{(r)})^\vee$. Therefore, we conclude that

\medskip

\begin{tabular}{@{$\bullet$ 6d theory of type\ }r@{\ on $S^1$ with\ }c@{\ twist gives the 5d\ }r@{\ theory.}}
$\SU(2\ell)$&  $\bZ_2$ & $\SO(2\ell+1)$\\[1ex]
$\SO(2\ell+2)$& $\bZ_2$ &  $\USp(2\ell)$\\[1ex]
$E_6$ & $\bZ_2$ &  $F_4$\\[1ex]
$\SO(8)$ & $\bZ_3$ &$G_2$
\end{tabular}

\medskip
\noindent Note that $\SO(2\ell+1)$ is \emph{not} a subgroup of $\SU(2\ell)$, and $\USp(2\ell)$ is \emph{not} a subgroup of $\SO(2\ell+2)$, either. This reiterates the fact that \emph{there is no gauge group $G$ in the 6d $\cN=(2,0)$ theory.} There are many attempts to write down the Lagrangian for 6d $\cN=(2,0)$ theory e.g.~\cite{Lambert:2010wm,Terashima:2010ji,Kawamoto:2011ab,Honma:2011br,Papageorgakis:2011xg,Ho:2011ni,Singh:2011id,Samtleben:2011fj,Chu:2011fd}, but correct ones i) should  be consistent only for $G=A,D,E$ \cite{Henningson:2004dh,Seiberg:2011dr}, and ii) should reproduce the behavior under twisted compactification listed above. 

This gives a  geometric explanation why the S-dual of 5d theory with non-simply-laced gauge group on $S^1$ is a simply-laced  theory twisted around $S^1$  by an outer automorphism. This also explains why the Seiberg-Witten curve of a 4d pure $\cN=2$ theory with non-simply-laced gauge group is given in terms of the spectral curve of the twisted simply-laced Toda system \cite{Martinec:1995by}.
Furthermore, twists by more general $\sigma_r(s_0,\ldots,s_\ell)$ can be incorporated by turning on $C_{(4)}$ of the Type IIB theory, or equivalently the flat background of the self-dual two-form fields $B_{(2)}^i$ of the 6d theory.

However, we have one difficulty. The analysis of this section, applied to the 6d theory of type $\SU(2\ell+1)$ with $\bZ_2$ twist line, seems to suggest that 5d $\SU(2\ell+1)$ theory with $\bZ_2$ twist around $S^1$ is S-dual to 5d  theory without a twist for a certain group. But in Sec.~\ref{affineLanglands}, we saw the  5d $\SU(2\ell+1)$ theory on $S^1$ with $\bZ_2$ outer-automorphism twist is self-dual. These two statements are clearly contradictory. 
We will claim in Sec.~\ref{partII} that this discrepancy is due to various subtle $-1$ signs which only comes with the $\bZ_2$ twist line of the 6d theory of type $\SU(2\ell+1)$.
But before that, we need to study the spectrum of dyons.

\section{Dyons}\label{dyons}
\subsection{Semi-classical quantization and half-BPS mass formula}\label{quantization}
So far our analysis was purely classical; let us perform the semi-classical quantization and consider the spectrum of dyons. For the quantization of monopoles in 4d $\cN=4$ theory, see e.g.~\cite{Osborn:1979tq,Sen:1994yi,Gauntlett:1996cw,Dorey:1996jh,Lee:1996if,Fraser:1997nd}. For pedagogical reviews, see \cite{Harvey:1996ur,Weinberg:2006rq}.

Consider the monopole-string solution \eqref{config1}, \eqref{config2} of charge $\aff\alpha^\vee=(k,\alpha)^\vee$.
It has a zero mode given by \begin{equation}
\Phi(s)=e^{is \alpha^\vee /2} \Phi e^{-is \alpha^\vee/2 },\qquad
A_5(s)=e^{is \alpha^\vee/2 } A_5 e^{-is \alpha^\vee/2 }\label{zeromode}
\end{equation} where we need to identify $s\sim s+2\pi$. 
Semiclassical states are then given by a wavefunction $\psi(s)$ on the space of $s$.\footnote{There is also the center-of-mass mode. In addition, it is known that there are even more zero modes when $(k,\alpha)$ is outside of the affine Weyl chamber defined by $\vev{\Phi}$. Careful treatment of these additional modes are outside of the scope of this paper.} 
The theta angles of the system affect  the boundary condition so that $\psi(s+2\pi)=\exp(i\theta_{\aff\alpha^\vee})\psi(s)$.
Then the wave function is of the form $\psi(s)=e^{in's}$ where $n'=n+\theta_{\aff\alpha^\vee}/(2\pi)$, for an integer $n$.
We will determine $\theta_{\aff\alpha^\vee}$ momentarily. For now let us consider it given.

The electric charge of the state is determined by the phase of the wavefunction under the global gauge transformation; the KK charge of the state is determined by the phase of the wavefunction under the shift of $x_5$.   
Note that the global gauge transformation by $e^{ih}$ for a Cartan element $h\in \fh$ shifts $s$ by $\alpha\cdot h$. Also, the shift $x_5\to x_5+c$ also corresponds to the shift of $s$ by $2\pi ck/R_5$.
Therefore, the electric charge is $n'\alpha$ and the KK charge is $n'k$, and the charge of the dyon obtained by quantizing the monopole of charge $(k,\alpha)^\vee$ is summarized as \begin{equation}
(k'_e,\alpha'_e;k_m,\alpha_m)=((n+\theta_{\aff\alpha^\vee})\aff\alpha ; \aff\alpha^\vee)
\end{equation} where $k'_e,\alpha'_e,k_m,\alpha_m$ are the KK, the electric, the instanton, and the magnetic charges, respectively. 

Given the validity of Sen's conjecture \cite{Sen:1994yi} for 4d $\cN=4$ $\SU(2)$ theory, there are states of charges \begin{equation}
(k'_e,\alpha'_e;k_m,\alpha_m)=((n+m\theta_{\aff\alpha^\vee})\aff\alpha ; m\aff\alpha^\vee)
\end{equation} for coprime integers $(n,m)$.

The mass of the state is given by rescaling that of the 4d $\SU(2)$ theory, and is \begin{equation}
m_\text{dyon}=|in+\frac{4\pi R_5 m}{g^2}\frac{2}{\alpha\cdot\alpha} | \cdot |\aff\alpha\cdot\aff\Phi|.
\end{equation}
This mass formula, deduced from embedding a know solution, is a special case of the general BPS mass formula.  Let us turn on a vev $\vev{\Phi_a}$ which is in a general direction of five scalars, $a=1,2,3,4,5$.  Consider a state with charge $(k'_e,\alpha'_e;k_m,\alpha_m)$.
The central charge $Z$ of a 4d $\cN=4$ system is a complex six-component vector, which is given by \begin{equation}
Z_{k_e,\alpha_e;k_m,\alpha_m}=(k'_e + i\frac{4\pi R_5}{g^2} k_m; (\alpha'_e + i\frac{4\pi R_5}{g^2}\alpha_m)\cdot \vev{\Phi_a}_{a=1,2,3,4,5}).
\end{equation}   Then the mass is given by $|Z|$.
Note that in this form $k'_e$ and $\alpha'_e$ are in general non-integral.

The central charge $Z$ of a half-BPS state should satisfy $\Re Z \propto \Im Z$ as six-component vectors. This means, in general, the charges of half-BPS states are of the form \begin{equation}
(k_e,\alpha_e;k_m,\alpha_m)=(p'(k,\alpha);q(k,\alpha)) 
\end{equation} and its mass is given by \begin{equation}
|Z|= |p' + i\frac{4\pi R_5}{g^2} q| \, |\aff\alpha\cdot\aff \Phi|.
\end{equation} Again, $p'$ is in general non-integral. If $p'$ is uniformly of the form $p'=p+\theta/(2\pi)$ for an integer $p$ and an angle $\theta$, this can be written as \begin{equation}
|Z|= |p + \tau q| \, |\aff\alpha\cdot\aff \Phi| \quad\text{where}\quad
\tau=i\frac{4\pi R_5}{g^2}+\frac{\theta}{2\pi}.\label{qeqm}
\end{equation} By choosing $\tau$ appropriately, we can make particles with a fixed ratio of $p:q$ lighter than all the other particles. They can be thought of as the electric particles in the appropriate duality frame.

Let us determine the phase shift $\theta_{\aff\alpha^\vee}$.
This is affected by two types of theta angles in our system. One is the discrete 5d theta angle $\theta_{5d}=0,\pi$ for the gauge group $\USp(2\ell)$, which measures the topology $\pi_4(\USp(2\ell))=\bZ_2$. For notational simplicity, we set $\theta_{5d}=0$ for other gauge groups.
Another comes from the background $\U(1)$ gauge potential $C$ which couples to the 5d super Yang-Mills via \begin{equation}
\propto \int C\wedge \Tr F\wedge F.\label{4dtheta}
\end{equation}  We let the curvature of $C$ to be zero. 
We normalize the coupling so that a single BPST instanton on the spacetime embedded  into $G$ via the $\SU(2)_\alpha$ couples to $C$ with the coefficient $i(\alpha^\vee\cdot\alpha^\vee/2)\int C$ in the Euclidean path integral. 
We let  $\theta_{4d}=\int_{S^1} C$, which plays the role of the 4d theta angle. 
$\theta_{4d}$ has period $2\pi$ for an untwisted gauge theory when we normalize $\alpha\cdot\alpha=2$.

The phase shift $\theta_{\aff\alpha^\vee}$ is  the phase assigned in the path integral to an adiabatic spacetime configuration on $S^1_t \times \bR^3 \times S^1_{x^5}$ where $s$ slowly changes from $0$ to $2\pi$ when going around $S^1_t$.
This configuration, considered on $S^1_t\times \bR^3$, has instanton number $2/(\alpha\cdot\alpha)$, and therefore the phase from \eqref{4dtheta} is $\theta_{4d}$. In addition, as is shown in Appendix~\ref{A}, the gauge configuration for $(k,\alpha)$ and that for $(k+1,\alpha)$ for a long root $\alpha$ of  $G=\USp(2\ell)$ differ by a nontrivial element of $\pi_4(\USp(2\ell))$, and thus we have \begin{equation}
\theta_{(k,\alpha)^\vee}=\frac{\alpha^\vee\cdot\alpha^\vee}2\theta_{4d}\qquad
 \text{or}\  = \frac{\alpha^\vee\cdot\alpha^\vee}2\theta_{4d}+\pi k,
\end{equation} where the latter is when $G=\USp(2\ell)$, $\theta_{5d}=\pi$, and $\alpha$ is a long root.

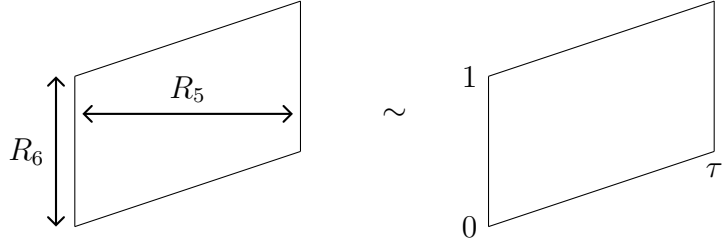
\begin{figure}\[
%\ncludegraphics[scale=.4]{slanted}
\begin{tikzpicture}
%	\begin{pgfonlayer}{nodelayer}
		\node  (0) at (0, 2) {};
		\node  (1) at (5.5, 2) {};
		\node  (2) at (-3.25, 1) {};
		\node  (3) at (-3, 1) {};
		\node  (4) at (2.25, 1) {$1$};
		\node  (5) at (2.5, 1) {};
		\node  (6) at (-2.9, 0.5) {};
		\node  (7) at (-0.1, 0.5) {};
		\node  (8) at (1.25, 0.5) {$\sim$};
		\node  (9) at (0, 0) {};
		\node  (10) at (5.5, 0) {};
		\node  (11) at (5.5, -0.25) {$\tau$};
		\node  (12) at (-3.25, -1) {};
		\node  (13) at (-3, -1) {};
		\node  (14) at (2.25, -1) {$0$};
		\node  (15) at (2.5, -1) {};
%	\end{pgfonlayer}
%	\begin{pgfonlayer}{edgelayer}
		\draw [<->,style=mine] (6.center) to node[above]{$R_5$} (7.center);
		\draw [<->,style=mine] (2.center) to node[left]{$R_6$} (12.center);
		\draw (1.center) to (10.center);
		\draw (3.center) to (0.center);
		\draw (15.center) to (5.center);
		\draw (5.center) to (1.center);
		\draw (13.center) to (3.center);
		\draw (9.center) to (13.center);
		\draw (0.center) to (9.center);
		\draw (10.center) to (15.center);
%	\end{pgfonlayer}
\end{tikzpicture}
\]
\caption{6d theory of type $G$, put on a slanted torus. The complex structure of the torus is identified with $\tau$ in \eqref{tau} \label{slanted}}
\end{figure}

For simply-laced theory on $S^1$ without outer-automorphism twist, $\theta_{\aff\alpha^\vee}=\theta_{4d}$, and the mass spectrum of dyons is then given by \begin{equation}
|Z|=|n+\tau m| \cdot |\aff\alpha\cdot\aff\Phi|\quad\text{where}\quad
\tau=i\frac{4\pi R_5}{g^2}+\frac{\theta_{4d} }{2\pi}.\label{tau}
\end{equation} which clearly shows the covariance under $\SL(2,\bZ)$ duality. 
Its realization in terms of the 6d theory is also straightforward: we identify the complex structure of the torus with $\tau$ just defined with $1/R_6=4\pi/g^2$, see Fig.~\ref{slanted}. 
Note that  $\theta_{4d}=\int_{S^1_{x^5}} C$ couples to the instanton charge, which is the KK momentum around the sixth direction. Therefore, the one-form $C$ is the part of the metric which skews the torus.

\subsection{$\SU(2\ell+1)$ with twist and $\USp(2\ell)$ with $\theta_{5d}=\pi$}\label{queer}
Let us apply the general formula obtained above to three cases:
\begin{itemize}
\item The first is the $\SU(2\ell+1)$ theory with the twist $\sigma_2(\ell)$ of order two, defined in \eqref{bar}.
Let us use the orthonormal basis $e_1$,\ldots, $e_{2\ell+1}$ with $\sum e_i=0$ as the Cartan subalgebra of $\SU(2\ell+1)$. Under the twist, $e_i$ is mapped to $-e_{2\ell+2-i}$. Therefore, the invariant part of the Cartan subalgebra can be described by $e_1$, \ldots, $e_\ell$.
The charge of the KK W-bosons is then given by \begin{equation}
(k_e,\alpha_e)=(k,\pm e_i \pm e_j),\
(k,\pm e_i),\ 
(k+\frac12,\pm 2e_i),\ 
(k+\frac12,\pm e_i\pm e_j),\ 
(k+\frac12,\pm e_i).
\label{X}
\end{equation} Here and in the following, we always use the convention that  $k$ is an integer, and $1\le i< j \le \ell$.

\item The second is the $\SU(2\ell+1)$ theory with the twist $\sigma_2(0)$ of order four, defined in \eqref{bar}.
Using the same basis of the invariant part of the Cartan, we have the spectrum of KK W-bosons given by \begin{equation}
(k_e,\alpha_e)=(k,\pm e_i \pm e_j),\
(k,\pm 2e_i),\ 
(k\pm\frac14,\pm e_i),\ 
(k+\frac12,\pm e_i\pm e_j)
\label{Y}
\end{equation} for $k\in \bZ$, $i\ne j$.

\item The third is the $\USp(2\ell+1)$ theory with $\theta_{5d}=\pi$. The charges of the KK W-bosons are not affected by the discrete theta angle, and is given by \begin{equation}
(k_e,\alpha_e)=(k,\pm 2e_i),\ (k,\pm e_i\pm e_j)
\label{Z}
\end{equation} for $k\in \bZ$, $i\ne j$.
\end{itemize}
We are going to check in the following three subsections that these spectra are permuted under $\SL(2,\bZ)$ duality action
as follows: \begin{equation}
\begin{array}{ll||c|c|c}
\text{Theory}&\setminus\text{States with} & p:q=1:0  & p:q=0:1 & p:q=1:1 \\ 
\hline
 \text{$\SU(2\ell)+1$} & \text{with twist $\sigma_2(\ell)$} & \eqref{X} & \eqref{Y} & \eqref{Z} \\
 \text{$\SU(2\ell)+1$}& \text{with twist $\sigma_2(0)$} & \eqref{Y} & \eqref{X} & \eqref{Z} \\
 \text{$\USp(2\ell)$}&\text{with  $\theta_{5d}=\pi$} & \eqref{Z} & \eqref{X} &\eqref{Y}
\end{array} \label{table}
\end{equation}
The second and the third columns can be interchanged by shifting $\theta_{4d}$.

\subsubsection{$\SU(2\ell+1)$ with the twist $\sigma_2(\ell)$ }\label{5.2.1}
The charges $(k_e,\alpha_e;k_m,\alpha_m)$ of the dyons are given by 
\begin{alignat}{3}
(n(k, \pm e_i \pm e_j ) &;  m (k,\pm e_i\pm e_j) ),\qquad &(n(k+\frac12, \pm e_i \pm e_j ) &;  m (k+\frac12,\pm e_i\pm e_j) )   \label{AA}\\
(n(k, \pm e_i  ) &;  m (2k,\pm 2e_i) ),\qquad  & (n(k+\frac12, \pm e_i  ) &;  m (2k+1,\pm 2e_i) ) \label{CC}\\
(n(k+\frac12, \pm 2e_i  ) &;  m (\frac{k}2+\frac14,\pm e_i) ) \label{EE}.
\end{alignat} Here, $(n,m)$ are coprime integers.

The states with $p:q=1:0$ are the ones with $(m,n)=(1,0)$, and give  the spectrum  \eqref{X} of the KK bosons of $\SU(2\ell+1)$ theory with twist $\sigma_2(\ell)$.
The states with $p:q=0:1$ are the ones with $(m,n)=(0,1)$, and give the spectrum  \eqref{Y} of the KK bosons of $\SU(2\ell+1)$ theory with twist $\sigma_2(0)$.
 The states with $p:q=1:1$ are the ones with $(m,n)=(1,1)$ in \eqref{AA}:
\begin{equation}
(k,\pm a_i \pm a_j), \qquad
(k+\frac12,\pm a_i \pm a_j),
\end{equation}
the ones with $(m,n)=(2,1)$ in \eqref{CC}: \begin{equation}
(k,\pm 2 a_i ),
\end{equation} and the ones with $(m,n)=(1,2)$ in \eqref{EE}: \begin{equation}
(k+\frac12,\pm 2 a_i ).
\end{equation} Together, they give  the spectrum  \eqref{Z} of the KK bosons of $\USp(2\ell)$ theory, rescaled by 2.

\subsubsection{$\SU(2\ell+1)$ with the twist $\sigma_2(0)$ }\label{5.2.2}
Next, let us consider dyons of $\SU(2\ell+1)$ theory with the twist $\sigma_2(0)$ of order four.
The charge spectra are given by \eqref{AA}--\eqref{EE} with the left hand side and the right hand side of the semicolon interchanged. Therefore, we easily conclude i) that 
the states with $p:q=1:0$ give the spectrum \eqref{Y} of the KK bosons of $\SU(2\ell+1)$ theory with twist $\sigma_2(0)$, ii) that 
the states with $p:q=0:1$ give the spectrum  \eqref{X} of the KK bosons of $\SU(2\ell+1)$ theory with twist $\sigma_2(\ell)$, and iii) that
the states with $p:q=1:1$  give  the spectrum  \eqref{Z} of the KK bosons of $\USp(2\ell)$ theory, rescaled by 2.

\subsubsection{$\USp(2\ell+1)$ with $\theta_{5d}=\pi$}\label{5.2.3}
Finally, let us consider $\USp(2\ell)$ theory with $\theta_{5d}=\pi$. The charges $(k_e,\alpha_e;k_m,\alpha_m)$ of dyons are given by \begin{align}
(n(k,\pm e_i\pm e_j)&; m(k,\pm e_i\pm e_j)), \label{XX} \\
((n-\frac{mk}2) (k,\pm 2e_i)&; m(k,\pm 2e_i). \label{YY}
\end{align} Note that the shift $mk/2$ in \eqref{YY} is due to the discrete theta angle. 
States with $p:q=1:0$ are given by those with $m=0$, which are the spectrum \eqref{Z} of the KK bosons of $\USp(2\ell)$ theory. 

States with $p:q=0:1$ arise from taking $n=0$ in \eqref{XX} giving \begin{equation}
(k,\pm e_i\pm e_j),
\end{equation}
and taking $(n,m,k)=(0,1,\text{even})$ or $(n,m,k)=(1,2,\text{odd})$ in \eqref{YY} giving respectively \begin{equation}
(k,\pm e_i), \quad (2k+1,\pm 2e_i).
\end{equation} Together, they form the spectrum \eqref{X} of the KK bosons of $\SU(2\ell+1)$ theory with twist $\sigma_2(\ell)$, scaled by a factor of 2.

States with $p:q=1:1$ arise from taking $(n,m)=(1,1)$ in \eqref{XX}  giving \begin{equation}
(k,\pm e_i\pm e_j),
\end{equation}
and taking $(n,m,k)=(0,1,\text{odd})$ or $(n,m,k)=(1,2,\text{even})$ in \eqref{YY} giving respectively \begin{equation}
(k+\frac12,\pm e_i), \quad (2k,\pm 2e_i).
\end{equation} Together, they form the spectrum \eqref{Y} of the KK bosons of $\SU(2\ell+1)$ theory with twist $\sigma_2(0)$, scaled by a factor of 2. This completes the check of the table \eqref{table}.

\subsection{Analysis using D4-branes and T-duality}\label{hananytroost}
When $G$ is a classical gauge group, the spectrum of  half-BPS dyons can be studied also using D4-branes and T-duality, and the relation to the twisted affine Lie algebras was already pointed out in \cite{Hanany:2001iy}. 

As an example, consider Type IIA setup  on a flat 10d spacetime with coordinates $x^{0,\cdots,3}$ and $x^{5,\cdots,10}$,  $2\ell$ D4-branes extending along $x^{0,\cdots,5}$.
We let the $i$-th D4-branes be at generic points, $x^{6,\cdots,10}=x^{6,\cdots,10}_i$. 
We then compactify $x^5\sim x^5+2\pi$, and perform the orientifold $(x^{6,\cdots,10}) \to (-x^{6,\cdots,10})$, placing the $\top4$-plane there. 
In~\cite{Hori:1998iv}, it was speculated that this system realizes 5d $\USp(2\ell)$ theory with $\theta_{5d}=\bZ_2$.\footnote{Also see~e.g.~\cite{Gimon:1998be,Gukov:1999yn,Hanany:2000fq,deBoer:2001px,Bergman:2001rp} for the behaviors of $\top4$-plane.}  Here we will see that our analysis in Sec.~\ref{5.2.3} can be thought of as confirmation of this hitherto-unchecked statement. 

Indeed, taking the T-dual around $x^5$, we have Type IIB system with an  O$3^+$-plane at $\tilde x^5=0$ and an $\top3$-plane at $\tilde x^5=\pi$, and all the resulting D3-branes are at $\tilde x^5=0$. The mass of $(p,q)$-string connecting the D3-branes 
satisfy the mass formula \eqref{qeqm}, with $\tau$ identified with the Type IIB axiodilaton. 
In particular, the fundamental string gives KK W bosons, the D-string gives instantonic monopoles, and the (1,1)-string gives towers of (1,1) dyons, see e.g.~\cite{Landsteiner:1996kf} for an analysis for 4d $\cN=4$ theory.
As we know which $(p,q)$-string can end on which type of O3-planes, we can  find  their spectra, which reproduce the result we found in Sec.~\ref{5.2.3}.

The twisted $\SU(2\ell+1)$ theory we studied in Sec.~\ref{5.2.1} and Sec.~\ref{5.2.2} can be analyzed in a similar way, and in Type IIB setup after T-duality, they are just images under the Type IIB $\SL(2,\bZ)$ action of the O3$^+$-$\top3$ setup considered above, as shown e.g. in~\cite{Hori}.

\section{Interpretation via 6d $\cN=(2,0)$ theory, part II} \label{partII}
In this section, we use the spectra of dyons studied in the previous section to study the properties of 6d $\cN=(2,0)$ theory of type $G=\SU(2\ell+1)$ in the presence of $\bZ_2$ twist lines. 

The 6d theory of type $G=\SU(2\ell+1)$ can be realized in two ways: one is from $2\ell+1$ coincident M5-branes in M-theory, and the other is from type IIB superstring put on the singularity $\bC^2/\bZ_{2\ell+1}$. 
Going to the Coulomb branch corresponds to separating coincident M5-branes in the first description, and blowing up the singularity in the second description. In the second description, we have two-cycles \begin{equation}
D_1,D_2,\ldots, D_{2\ell}
\end{equation} corresponding to the nodes of the Dynkin diagram of type $\SU(2\ell+1)$, so that \begin{equation}
D_i\cdot D_j = 1 \ \text{if}\  j=i\pm 1, \quad D_i\cdot D_j=0\ \text{otherwise}.
\end{equation}
The nontrivial $\bZ_2$ outer automorphism of $\SU(2\ell+1)$ should map $D_i$ to $D_{\ell+1-i}$, but the hyperk\"ahler isometry $g$ on the blown-up $\bC^2/\bZ_{2\ell+1}$ mapping $D_i$ to $D_{\ell+1-i}$ generates the group $\bZ_4$, i.e. $g^2\ne 1$, $g^4=1$.\footnote{To see this, consider a neighborhood $(x,y)\in\bC^2$ of the intersection of $D_\ell$ and $D_{\ell+1}$, so that the defining equations of $D_{\ell}$ and $D_{\ell+1}$  are given by $x=0$ and $y=0$, respectively. Then the hyperk\"ahler isometry which exchanges them is $(x,y)\mapsto (-y,x)$, which is of order 4. }
Nevertheless, $g^2$ is believed to act trivially on the low-energy 6d $\cN=(2,0)$ theory.
Then there is a $\bZ_2$ action on the low-energy theory, which can be identified with the M-theory orientifold action studied in \cite{Hori:1998iv,Gimon:1998be,Hori}.

The peculiar feature of the $\bZ_2$ action on the 6d theory of type $\SU(2\ell+1)$ is that $D_\ell$ is mapped to $D_{\ell+1}$, and $D_\ell \cdot D_{\ell+1}=1$, see Fig.~\ref{outer}.  Note that in all other cases, namely the $\bZ_2$ action on $\SU(2\ell)$, $\SO(2\ell)$, $E_6$ and $\bZ_3$ action on $\SO(8)$, when $D_i$ is mapped to  a different cycle $D_{\sigma(i)}$ under the outer automorphism, the intersection number is always zero: $D_i\cdot D_{\sigma(i)}=0$.
This difference is reflected in the intersection form of the three-cycles of the twisted fibration of the blownup $\bC^2/\Gamma(G)$ over $T^2$: it is odd only when $G=\SU(2\ell+1)$ with $\bZ_2$ twist, and even otherwise.
This feature seems to produce various subtle minus signs in the 6d theory which the authors still do not quite understand.

\begin{figure}\[
\includegraphics[scale=.4]{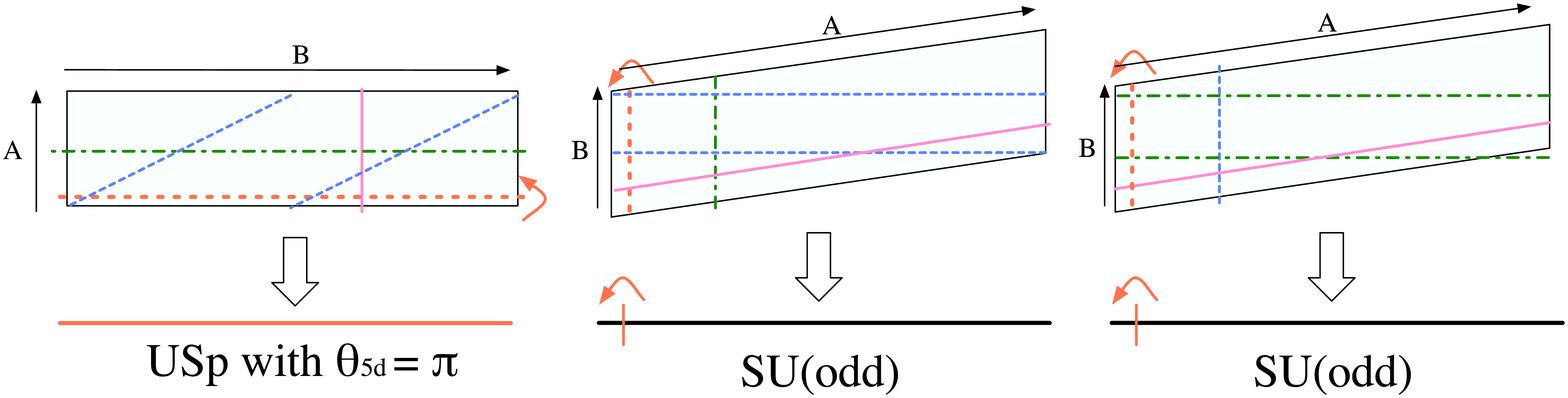}
\]
\caption{ 6d theory of type $\SU(2\ell+1)$ put on a torus with $\bZ_2$ twist, together with wrapped strings.  Orange broken lines are the $\bZ_2$ twist lines, across which the $\bZ_2$ operation is performed.  Strings wrapped around pink solid, blue dotted, and green chained lines produce particles with charges which match the KK W-bosons of 
$\USp(2\ell)$ theory, $\SU(2\ell+1)$ theory with twist $\sigma_2(\ell)$, $\SU(2\ell+1)$ theory with twist $\sigma_2(0)$, respectively.  \label{usp}}
\end{figure}

Consider the 6d theory of type $\SU(2\ell+1)$ put on  a torus with $\bZ_2$ twist line. We set the holonomy of the self-dual 2-forms on $T^2$ to be zero.
Let $A$ the 1-cycle around which there is a nontrivial $\bZ_2$ action, and $B$ the other 1-cycle around which there is no nontrivial $\bZ_2$ action, so that the intersection number is $A\cdot B=1$.
In Fig.~\ref{usp}, the $B$-cycle is the cycle parallel to the orange broken line,
and the $A$-cycle is the cycle intersecting it. 
Then the spectra of the dyons found in Sec.~\ref{queer} translates to the following properties: 
\begin{enumerate}
\item When the $A$-cycle is shorter, the system reduces to 5d $\USp(2\ell)$ theory with $\theta_{5d}=\pi$, see the leftmost figure of Fig.~\ref{usp}.
\item The quantization of the string wrapped on a one-cycle around which there is a nontrivial $\bZ_2$ action, i.e.~a cycle of the form $nA+mB$ for an odd $n$, gives rise to the spectrum \eqref{Z} of KK W-boson of $\USp(2\ell)$ theory. In Fig.~\ref{usp}, it is shown in solid pink lines. 
\item When the $B$-cycle is shorter, the system reduces to 5d $\SU(2\ell+1)$ gauge theory. This is a known property of the 6d theory of type $\SU(2\ell+1)$. The $\bZ_2$ twist line becomes an action of a $\bZ_2$ outer automorphism of the $\SU(2\ell+1)$ gauge group, see  the middle and the rightmost figures of Fig.~\ref{usp}. Even with zero background self-dual two-form, the precise automorphism can either be $\sigma_2(0)$ or $\sigma_2(\ell)$,  defined in \eqref{bar}.
\item The quantization of the string wrapped on a one-cycle around which there is no nontrivial $\bZ_2$ action, i.e.~a cycle of the form $nA+mB$ for an even $n$, gives rise to either the spectrum \eqref{X} of KK W-boson of $\SU(2\ell+1)$ theory with twist $\sigma_2(\ell)$, or the spectrum \eqref{Y} of KK W-boson of $\SU(2\ell+1)$ theory with twist $\sigma_2(0)$. When the quantization on a string wrapped on one such one-cycle $B'$ gives the spectrum \eqref{X}, then the quantization on a string wrapped on $B'+2A$ should give the spectrum \eqref{Y} and vice versa. In Fig.~\ref{usp}, the cycle producing the spectra \eqref{X} and \eqref{Y} are shown in blue dotted and green chained lines, respectively.
\item The gauge theory $\tau$ of the twisted $\SU(2\ell+1)$ theory, appearing say in the BPS mass formula, and the geometric complex structure $\tau_\text{geom}$ of the $T^2$ on which the 6d theory is compactified are related as \begin{equation}
\tau = 2\tau_\text{geom}+1.
\end{equation}
\end{enumerate}

\begin{figure}\[
\includegraphics[scale=.4]{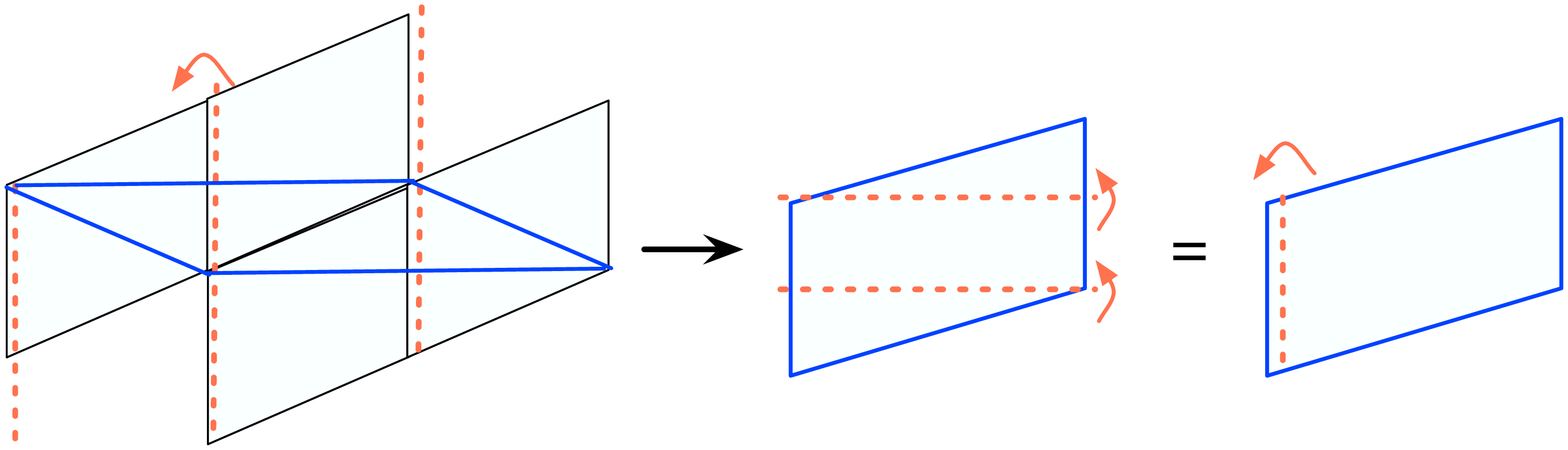}
\]
\caption{S-duality applied to 5d $\SU(2\ell+1)$ theory with $\bZ_2$ twist at $\theta_{4d}=0$. \label{slantedS}}
\end{figure}

A few comments are in order. 
\begin{itemize}\item 
As for the point 1), it was shown in \cite{Hori} that $2\ell+1$ M5-branes on an $S^1$, around which the M-theory orientifolding flipping all the transverse direction of M5-branes is performed, reduce in the Type IIA language to the system of $2\ell$ D4-branes on top of an $\top4$-plane. As recalled in Sec.~\ref{hananytroost}, this system was guessed to produce  5d $\USp(2\ell)$ theory with $\theta_{5d}=\pi$, and the property 1) is consistent with it. Our field-theoretic analysis in Sec.~\ref{5.2.3} supports this.

\item 
As for the point 2), the strings wrapped on the solid pink line correspond to D3-branes wrapped on the 3-cycle of the form $A\times(D_i+D_{2\ell+1-i})$, when we realize the system using type IIB on the blown-up $\bC^2/\bZ_{2\ell+1}$.  The roots of $\USp(2\ell)$ gauge bosons are of the form $\pm e_i\pm e_j$ for $i\ne j$ and $\pm 2 e_i$. Then the simple root $e_i-e_{i+1}$ corresponds to the cycle $(D_i-D_{i+1})+(D_{2\ell+1-i}-D_{2\ell-i})$, and $2e_{\ell}$ corresponds to $2(D_\ell + D_{\ell+1})$, as can be checked by calculating the intersections.  
This means that it is \emph{not allowed} to wrap the D3-brane on a cycle $A\times(D_\ell+D_{\ell+1})$, 
but \emph{is allowed and does produce a one-particle state} when the D3-brane is wrapped on a cycle $A\times 2(D_\ell+D_{\ell+1})$.

\item 
As for the points 3) and 4), it is instructive to compare with the twisted $G$ theory other than $\SU(2\ell+1)$ with $\bZ_2$ twist, treated in Sec.~\ref{partI}. Let us set the integral of the self-dual two-forms on $T^2$ to zero. 
Then, when $\tau$  is continuouly changed and becomes the value so that the system reduces to 5d $G$ theory with an outer automorphism twist, the twist was always $\sigma_r(0)$.

For $\SU(2\ell+1)$ theory with $\bZ_2$ twist line, however, the spectra of particles coming from strings wrapped on the cycle $B$ and on $B'=B+2A$ differ, although both $B$ and $B'$ have intersection $1$ with $A$ and have no nontrivial $\bZ_2$ action around themselves. Therefore, when $\tau_\text{geom}$ is changed continuoulsy and becomes the value so that the system reduces to 5d $\SU(2\ell+1)$ theory with an outer automorphism twist, the twist can either be $\sigma_2(\ell)$  or $\sigma_2(0)$, depending on the choice of $\tau_\text{geom}$.

\item 
The need for the different quantization of strings wrapped on cycles $B$ and $B+2A$ can also be understood as follows. Let us use the type IIB language, and consider the D3-brane wrapped on three-cycles $B\times D_\ell$ and $(B+2A)\times D_\ell$. They have intersection number 1, which means that the Dirac-Zwanziger pairing of these two particles is 1.  In $\cN=4$ 4d gauge theory, however, we never find  a pair of particles whose Dirac-Zwanziger pairing is 1. Therefore, the KK momenta assigned to the motion of the strings wrapped on $B$ and $B+2A$ need to be different, so that they do not simultaneouly have zero KK momenta to remain in the 4d spectrum.

\item 
As for the point 5), the difference in the gauge theory $\tau$ and the geometric $\tau_\text{geom}$ is necessary to reproduce the fact we saw in Sec.~\ref{affineLanglands} that $\SU(2\ell+1)$ theory with twist $\sigma_2(\ell)$ at $\tau=i4\pi /g^2$ is S-dual to $\SU(2\ell+1)$ theory with twist $\sigma_2(0)$ at $\tau=ig^2/4\pi$. 
The factor $2$ is basically a convention in the normalization of the gauge theory $\tau$; the shift by $1$ comes from the property that the gauge theory has $\theta_{4d}=0$ when $\theta_\text{geom}=\pi$; this is another manifestation of a mysterious $-1$.

Then the operation $\tau\to -1/\tau$ corresponds to the map \begin{equation}
\tau_\text{geom}\mapsto  \frac{\tau_\text{geom}-1}{2\tau_\text{geom}-1}
\end{equation} illustrated in Fig.~\ref{slantedS}, exchanging the middle and the rightmost figures of Fig.~\ref{usp}. 
\end{itemize}

The immediate problem is then to deduce the properties 1) -- 5) directly from the point of view of 6d $\cN=(2,0)$ theory, or utilizing the string theory realizations of it. The authors hope to come back to this problem in the near future.

\section*{Acknowledgements}
YT thanks P. Argyres, J. Distler, R. Eager, D. Gaiotto, K. Hori, G. W. Moore, D. R. Morrison, N. Seiberg, A. Shapere, S. Sugimoto and E. Witten for helpful discussions. 
He also thanks people who answered \href{http://tex.stackexchange.com/questions/5309/}{\texttt{http://tex.stackexchange.com/questions/5309/}} for replicating the look of the Dynkin diagrams found in \cite{Kac} by \LaTeX.
He is  currently supported in part by World Premier International Research Center Initiative (WPI Initiative),  MEXT, Japan through the Institute for the Physics and Mathematics of the Universe, the University of Tokyo.
He was working on this paper for almost two years, and most of the time he was in the Institute for Advanced Study, where he was  supported in part by NSF grant PHY-0969448  and by the Marvin L. Goldberger membership.

\appendix
\section{The effect of the discrete theta angle}\label{A}
In Sec.~\ref{quantization}, we needed the phase assigned to a field configuration on $S^1_t \times \bR^3\times S^1_{x^5}$ in our gauge theory path integral, and postponed the analysis of the $\USp(2\ell)$ gauge theory which has 5d discrete theta angle associated to $\pi_4(\USp(2\ell))$. Here we fill in the details. 

First, we consider only field configurations which vanish at the spatial infinity of $\bR^3$, together with the holonomy around $S^1_{x^5}$. Therefore, for our purposes, we can think of $\bR^3\times S^1_{x^5}$ as $S^4$.  We then need to understand the topology of the $\USp(2\ell)$ gauge bundles on $S^1 \times S^4$, where we assume $S^1$ refers to the time direction, unless otherwise noted. 
We identify a $\USp(2\ell)$ gauge configuration on $S^1\times S^4$ with the corresponding bundle $V$ of quaternion vector space over $S^1\times S^4$, which can also be represented as an element \begin{equation}
[V]-[\bH^\ell] \in \widetilde{\KSp}(S^1\times S^4).
\end{equation} Here and in the following, $[\bH^\ell]$ or $[\bR^k]$ stands for a trivial $\bH^\ell$ or $\bR^k$ bundle over the space, respectively. 
The right hand side can be decomposed using \begin{equation}
\widetilde{\KSp}(S^1\times S^4) = \widetilde{\KSp}(S^1\wedge S^4)\oplus \widetilde{\KSp}(S^1) \oplus \widetilde{\KSp}(S^4).
\end{equation} 
The 4d theta angle, which is a holonomy of a background one-form $\int_{S^1} C$, measures the component $\widetilde{\KSp}(S^4)\simeq \bZ$, % which is slightly wrong, because $\int_{S^1} C$ we used in Sec.~\ref{quantization} was around $S^1_{x^5}$. But whatever.
and the nontrivial 5d theta angle assigns a phase $-1$ to the nonzero element of $\widetilde{\KSp}(S^5)\simeq \pi_4(\USp(2\ell))\simeq \bZ_2$.  The factor $\widetilde{\KSp}(S^1)$ is trivial.

The generator of this $\bZ_2$ corresponds to the virtual bundle \begin{equation}
([\text{M\"obius}]-[\bR]) \otimes_\bR ([\text{BPST}]-[\bH]) \in \widetilde{\KO}(S^1)\otimes \widetilde{\KSp}(S^4) \to \widetilde{\KSp}(S^1\times S^4).
\end{equation} See e.g. \href{http://books.google.com/books?id=sYYgTtaf5rwC&pg=PA211}{Chapter IV, Theorem 5.16} of \cite{MimuraToda}.
Here [M\"obius] corresponds to the $\bR$-bundle over $S^1$ which is the M\"obius strip,
and [BPST] corresponds to the $\bH$-bundle over $S^4$ which is the standard BPST one-instanton solution of $\SU(2)$. Note that $[\text{M\"obius}]\otimes_\bR [\bH] = [\bR]\otimes_\bR [\bH]$ as an element of $\KSp(S^1\times S^4)$ because $-1$ is connected to the identify in $\USp(2)$, although not in $\mathrm{O}(1)$. Therefore \begin{equation}
[\text{M\"obius}]\otimes_\bR [\text{BPST}] \in {\KO}(S^1)\otimes {\KSp}(S^4) \label{qqq}
\end{equation} determines the generator of $\bZ_2$ under the natural map to $\widetilde{\KSp}(S^1\times S^4)$.

Now, the configuration on $S^1_t \times \bR^3 \times S^1_{x^5}$ we needed was the monopole-string solution of charge $(k,\alpha)^\vee$  for $k=1$ and $\alpha$ a long root when restricted to $\bR^3\times S^1_{x^5}$,  and around $S^1_t$ we performed the zero-mode rotation by $\alpha$ precisely once, see \eqref{zeromode}. 
Under the identification $\bR^3\times S^1_{x^5}\simeq S^4$, the bundle on $S^4$ has instanton number 1.
The zero-mode rotation does $-1\in \SU(2)\subset \USp(2\ell)$ for the $\SU(2)$ subgroup  specified by $\alpha$. Therefore, this 5d configuration has the topological type \begin{equation}
([\text{M\"obius}]\otimes_\bR [\text{BPST}]) \oplus [\bH^{2(\ell-1)}] \in \KO(S^1)\otimes \KSp(S^4)\to \KSp(S^4\times S^1)
\end{equation} which is equivalent to the element \eqref{qqq} as an element in $\widetilde{\KSp}(S^1\times S^4)$. Therefore when $\theta_{5d}=\pi$ for the $\USp(2\ell)$ theory on 5d, a minus sign is assigned to this field configuration. This was what was needed in Sec.~\ref{quantization}.

\bibliographystyle{ytphys}
%\small\baselineskip=.9\baselineskip\let\bbb\bibitem\def\bibitem{\itemsep1.5pt\bbb}
\bibliography{ref}
\end{document}